\documentclass[12pt,preprint]{aastex}
\usepackage{amsmath,longtable,natbib}
\usepackage{lineno}
\usepackage{float}
\usepackage[caption = false]{subfig}
\usepackage{hyperref}
\usepackage{graphicx}
\shorttitle{GHRSS $-$ update on the survey}
\shortauthors{Bhattacharyya et al.}
\usepackage{subfig}

\begin{document}

\title{Post-correlation beamformer for time-domain studies of pulsars and transients}
\author{
Jayanta~Roy\altaffilmark{1},
Jayaram N.~Chengalur\altaffilmark{1},
Ue-Li~Pen\altaffilmark{{2}{3}{4}{5}}}
\altaffiltext{1}{National Centre for Radio Astrophysics, Tata Institute of Fundamental Research, Pune 411 007, India}
\altaffiltext{2}{Canadian Institute for Theoretical Astrophysics (CITA), 60 St. George Street, Toronto, Canada M5S 3H8}
\altaffiltext{3}{Dunlap Institute for Astronomy and Astrophysics, University of Toronto, Toronto, ON M5S 3H4, Canada}
\altaffiltext{4}{Canadian Institute for Advanced Research, CIFAR Program in Gravitation and Cosmology, Toronto, ON M5G 1Z8, Canada}
\altaffiltext{5}{Perimeter Institute for Theoretical Physics, Waterloo, ON N2L 2Y5, Canada}
\affil{}

\begin{abstract}

We present a detailed analysis of post-correlation beamforming (i.e. beamforming which involves only phased sums 
of the correlation of the voltages of different antennas in an array), and compare it with the traditionally 
used incoherent and phased beamforming techniques. Using data from the GMRT we show that post-correlation 
beamformation results in a many-folds increase in the signal-to-noise for periodic signals from pulsars and several order of
magnitude reduction in the number of false triggers from single pulse events like fast radio bursts (FRBs). 
This difference arises primarily because the post-correlation beam contains less red-noise, as well as 
less radio frequency interference. The post-correlation beam can also be more easily calibrated than the incoherent 
or phased array beams. We also discuss two different modes of post-correlation beamformation, viz. (1) by subtracting 
the incoherent beam from the coherent beam and (2) by phased addition of the visibilities. The computational costs 
for both these beamformation techniques as well as their suitability for studies of pulsars and FRBs are discussed. 
Techniques discussed here would be of interest for all upcoming surveys with interferometric arrays. Finally, 
we describe a time-domain survey with the GMRT using the post-correlation beamformation as a case study. 
We find that post-correlation beamforming will improve the current GMRT time-domain survey sensitivity by 
$\sim$ 2 times for pulsars with periods of few 100s of millisecond and by many-folds for even slower pulsars, 
making it one of the most sensitive surveys for pulsars and FRBs at low and mid radio frequencies.

\end{abstract}

\vskip 0.6 cm

\section{Introduction}
\label{sec:intro}

Despite over five decades of pulsar surveys there are only $\sim$ 2600 pulsars that have been discovered 
so far\footnote{\url{http://www.atnf.csiro.au/people/pulsar/psrcat/}}. This is 5\% or less of the total 
Galactic population of pulsars (estimates of the Galactic population range from 40,000 to 90,000 objects, 
see e.g. \cite{Lorimer08}) and a large population of pulsars remains to be discovered by current and future 
surveys. Even though there has been an accelerating rate of discovery over the last decade, this has not 
been uniform across the entire parameter space occupied by pulsars. For example, even though the population 
of known Galactic field millisecond pulsars (MSPs) has increased approximately 4 fold over the last  decade\footnote{\url{http://astro.phys.wvu.edu/GalacticMSPs/GalacticMSPs.txt}}, there has only been a 
$\sim$ 40\% increase in the number of known slow pulsars (i.e. pulsars with period, P $>$ 30~ms). This very modest increase in the number of known slow pulsars is particularly unfortunate, since the already known population of relatively slow pulsars contains several 
interesting objects, such as double neutron stars ($\sim$ 15 known, \cite{Tauris17}), which enable tests of strong field gravity, energetic young pulsars with significant 
spin-down noise, normal pulsars showing intermittency, drifting, and nulling, probing hitherto unknown emission physics,
magnetars with extraordinary high magnetic fields  ($\sim$ 
29 known\footnote{\url{http://www.physics.mcgill.ca/~pulsar/magnetar/main.html}}), ultra slow pulsars 
(only 2 known) with period $>$ 10s which graze the theoretical death-line. One of the major reasons 
for the relatively slow increase in the number of such known pulsars is that the detection of these 
objects via periodicity searches can be severely affected by both instrumental red-noise and 
radio frequency interference (RFI). Both of these phenomena particularly reduce the search sensitivity at 
the low frequency end of the power spectrum of the detected time series, which is where the signal from these objects is strongest.

In addition to pulsars, the population of time-domain radio transients consists of Rotating Radio Transients 
(\cite{McLaughlin06}; 112 known\footnote{\url{http://astro.phys.wvu.edu/rratalog/}}) and Fast Radio Bursts 
(\cite{Lorimer07}, \cite{Thornton13}; 33 known\footnote{\url{http://www.frbcat.org/}}). All the 
Fast Radio Bursts (FRBs) discovered to date are single events (except for one repeating FRB) of 
millisecond duration with dispersion measure (DM) values generally higher than the possible Galactic contribution.
The non-repeating nature of these sources warrants real-time time-domain detections aided by simultaneous 
millisecond time-scale imaging to localize these events in order to maximise the science returns. 
Rotating Radio Transients (RRATs) show occasional flashes of dispersed radio bursts of typically a few 
milliseconds duration. The cause of their sporadic emission as well as their connection to other neutron star 
populations are not fully understood. Detection of a large number of FRBs and RRATs is essential in order for us 
to gain a better understanding of the  nature of these sources. However, detection of such single pulse events 
with millisecond duration in dedispersed time-series data is severely hindered by the presence of RFI.

Time-domain surveys are generally sensitivity limited, hence surveys with more sensitive instruments should lead to 
a higher discovery rate. Many of the existing as well as future high sensitivity radio telescopes are interferometric arrays. 
Planned surveys with telescopes like MeerKAT (e.g. TRAPUM\footnote{\url{http://www.trapum.org/}}), 
SKA Phase1 \citep[e.g.][]{Levin17} also need to optimally combine signals from many small telescopes 
(i.e. do ``beamformation''). The GMRT was one of the first interferometric instruments to be systematically 
used for pulsar searches. The high discovery rate of the GMRT High Resolution Southern Sky (GHRSS\footnote{\url{http://www.ncra.tifr.res.in/ncra/research/research-at-ncra-tifr/research-areas/pulsarSurveys/GHRSS}}; 
\cite{Bhattacharyya16,Bhattacharyya17}) as well as the Fermi-directed survey \citep{Bhattacharyya13}  demonstrate the capabilities of the GMRT for low frequency pulsar searches.
The recent upgrade of the GMRT allowing much larger instantaneous bandwidths (uGMRT; \cite{Gupta17}) 
brings a significant increase in its theoretical survey sensitivity for pulsars and FRBs at low and mid radio frequencies. 
With the uGMRT, phase-2 of the GHRSS survey \citep{Roy18} is expected to achieve a sensitivity better than all existing 
and ongoing off-galactic plane surveys. Most of the existing and planned surveys however use one of the 
two traditionally used methods of beamformation, viz. Incoherent Array (IA) or Phased Array (PA) beams, 
which are described in more detail below. In this paper we explore the possibility of significantly improving 
the observed time-domain sensitivity using yet another kind of beamformation, viz. ``post-correlation beamforming''. 
We show that in this kind of beamforming, the contribution of instrumental red-noise to the power spectrum is 
significantly reduced, thus greatly improving the sensitivity towards low and mid spin frequency pulsars. 
We also show that post-correlation beamformation can be used to significantly reduce the effect of RFI, 
thus improving the time-domain sensitivity for periodicity and single pulse search. Both of these factors lead to reduction 
of the number of false detections by several orders of magnitude. This not only allows one to lower the candidate 
detection threshold (i.e. probe fainter flux levels) but also greatly eases the problem of carrying out on-the-fly imaging 
and other follow-up of these events to maximize the science returns.

\section{Beamformation with Antenna Arrays}
\subsection{Incoherent Array and Phased Array beamformation}
\label{sec:post}

Two commonly used beamformation techniques for antenna arrays are Incoherent Array (IA) beamformation and Phased Array (PA) beamformation. For example, the GMRT backends (GSB; \cite{Roy10} or GWB; \cite{Reddy17}) allow one to form both these type of beams by making per spectral channel combinations of the delay and fringe corrected signals from different antennas.  The Incoherent Array (IA) beam is formed by summing together the 
squares of the individual antenna voltages, i.e. it adds together the  signal powers. Mathematically
\begin{equation}
  P_{\rm IA} = \sum_{i=0}^{N-1} |V{_i}|^{2}
\end{equation}
where $P_{\rm IA}$ is the IA beam signal and $V_i$ are the voltages from the individual antennas. This kind of 
combination leads to a wide field-of-view (but at reduced sensitivity compared to a phased combination) and 
is useful for blind searches (such as, for e.g. the GHRSS survey). The coherent or Phased Array (PA) beam is 
produced by summing together the voltages (after phasing them appropriately so that the beam points to the 
direction of interest) and then squaring the resultant sum. Mathematically,
\begin{equation}
  P_{\rm PA} = |\sum_{i=0}^{N-1} V{_i} e^{-i\phi_i}|^2
\end{equation}

where $P_{\rm PA}$ is the PA beam signal and $V_i$ are the delay and fringe corrected voltages from the 
individual antennas, and $\phi_i$ is the phase introduced in antenna $i$ in order to steer the beam towards the 
desired direction.  The PA beam has higher sensitivity than the IA beam. The signal-to-noise ratio ($S/N$) 
for observations of a single pulse of flux density S located at the pointing centre of a dual polarized array
for the IA and PA beam are:

\begin{equation}  
         (S/N)_{\rm IA}=\frac{GS \sqrt{2 N_{\rm a} \Delta{\nu} \tau} }{T_{\rm sys}} 
\label{eqn1}
\end{equation}

\begin{equation}
         (S/N)_{PA} =\frac{GS N_{\rm a}\sqrt{2 \Delta{\nu} \tau} }{T_{\rm sys}}
\label{eqn2}
\end{equation}

where G is the gain of a single telescope, $N_{\rm a}$ is the number of antennas used for beamformation, 
$\Delta{\nu}$ is the instantaneous observing bandwidth, $\tau$ is the integration time and $T_{\rm sys}$ 
is the total system noise. These expressions assume that the sky noise is small compared to the receiver 
noise of the antennas. The sensitivities of the IA and PA beams under different scenarios are discussed 
in detail in \cite{Kudale17}.

The IA beam is not only less sensitive that the PA beam, it is also more vulnerable to instrumental gain 
fluctuations and RFI. This is because the IA beam is the sum of the auto-correlations of the individual
antennas. Since most of the terms in the PA beam correspond to cross-correlations between antennas,
it has some immunity to RFI (which gets decorrelated by the delay tracking/fringe rotation operations)
as well as to fluctuations in the instrumental gains.
We illustrate this by showing in Fig.~\ref{fig:rednoise} the dedispersed time-series 
for PSR J2144$-$3933 from simultaneous IA and PA observations using the GMRT. As can be seen, fewer 
RFI bursts are seen in the PA beam as compared to the IA beam. The PA beam noise properties in general 
appear better than those of the IA beam; one can see individual single pulses in the de-dispersed 
{\it PRESTO} \citep{{Ransom02}} output, while these pulses are lost in the noise of the IA beam. 
Still further improvement in the noise properties can be seen in the post-correlation beam output 
(the lowest panel in the figure). We discuss this in more detail below.

\subsection{Post-Correlation Beamformation}
\label{sec:postcorr}

Post-correlation (PC) beamformation (e.g. \cite{Kudale17}), conceptually consists of forming the 
desired beam not by combining the individual antenna voltages, but rather by combining the 
(suitably phased) visibilities from the different baselines in the array. Effectively, this 
eliminates the auto-correlation terms from the PA beam. According to the radiometer equation 
(\ref{eqn1} and \ref{eqn2}), for an array with $N_{a}$ elements, in situations where sky noise is negligible (i.e. T$_{sky}$ $<<$ $T_{rec}$), 
the IA beam sensitivity scales as $\sqrt{N_{a}}$, whereas PA beam sensitivity scales as  $N_{a}$. Following equation-29 
of \cite{Kudale17}, post-correlation beam sensitivity scales as $\sqrt{N_{a}(N_{a}-1)}$. For a 
telescope like the GMRT with $N_{a}$ $=$ 30, the theoretical degradation of the sensitivity for the post-correlation beam
compared to the PA beam is $<$ 2\%. The reduction in sensitivity arises from the non inclusion of the $N_{a}$ auto-correlation terms. 
However, in practice, since the auto-correlation signals are the ones which are most affected by instrumental 
gain fluctuations and RFI, one could in fact (as can be seen in Fig.~\ref{fig:rednoise}) get a significant 
improvement in the signal to noise ratio when using PC beamformation instead of PA beamformation. In addition, 
the PC beam is also easier to calibrate. Basically, as far as calibration is concerned, since the PC beam 
consists only of visibility data, and assuming that visibilities are also computed in parallel 
(as is the case for the GMRT) the beam can be calibrated using exactly the same techniques as standard 
interferometric imaging calibration. This holds also for polarimetric calibration. Since both the 
IA and the PA beams contain auto-correlation terms, proper calibration of these beams involves 
calibration of the system temperature. We note that calibration of the post-correlation beam
could be done in real time, in situations where the visibility data is also output. For example,
\cite{Kudale17}) demonstrate for the GMRT that it is possible to apply in real time phases
obtained via in-field self calibration to keep the PA beam phased.
 
\begin{figure}
        \includegraphics[width=5in,angle=0]{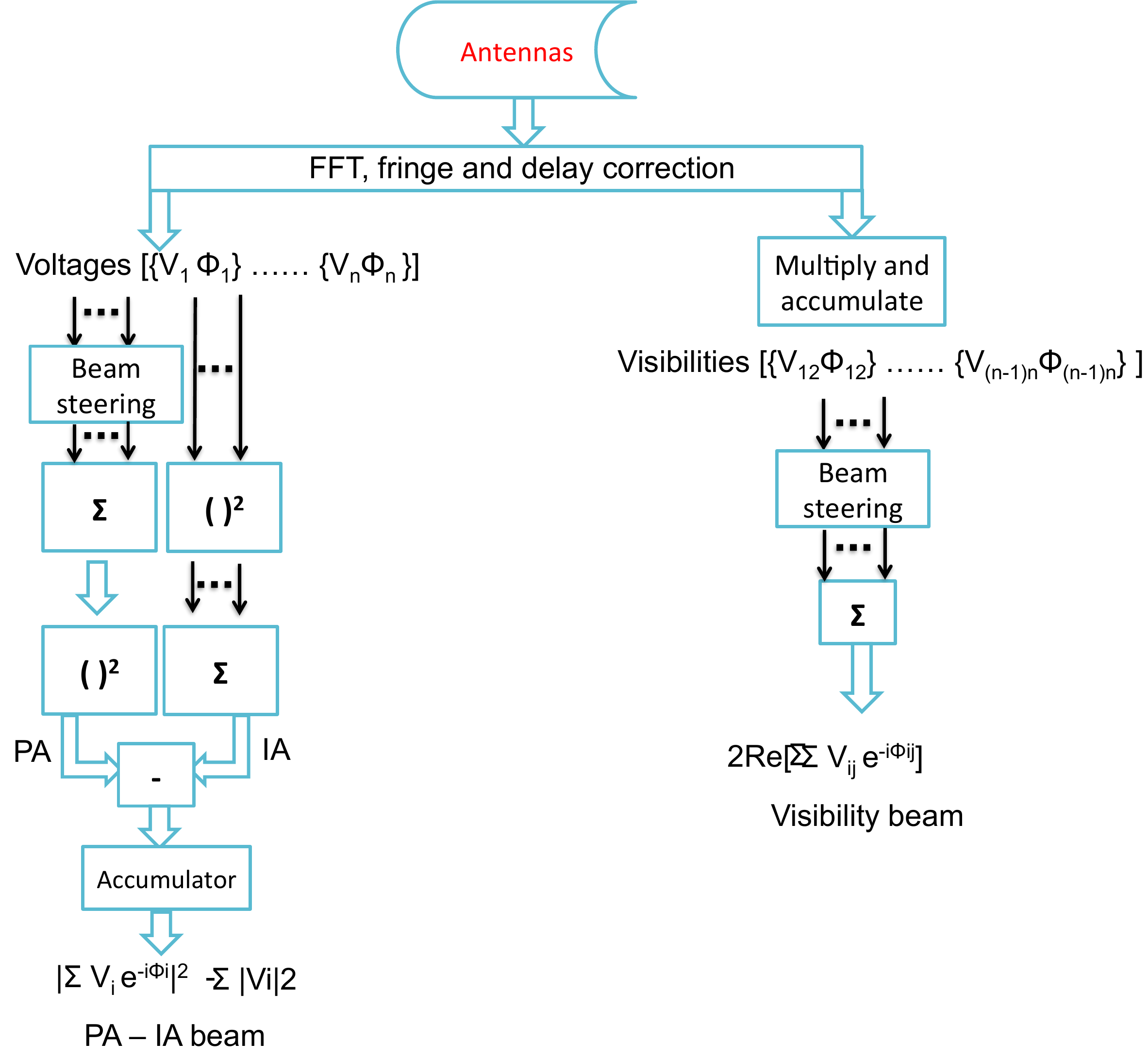}
        \caption{The schematic block diagram for post-correlation beamformation. The left branch shows PA-IA beamformation and the right
branch shows beamformation using visibilities. Note that in PA-IA beamformation the beam steering has to be done at the FFT block level,
while in visibility based beamformation the beam steering is done after accumulation.}
        \label{fig:PA-IA_block}
\end{figure}

Although the name implies that post-correlation beamformation has to be done after correlating 
the antenna voltages, the beam can in fact be operationally produced in two different ways; 
(1) by subtracting the IA beam from the PA beam; this effectively removes all the auto-correlation 
data that is contained in the PA beam, or (2) by a phased addition of the cross visibilities. 
Mathematically, we could do either of

\begin{equation}
  P_{\rm PC} = |\sum_{i=0}^{N} V_{i} e^{-i\phi_{\rm i}}|^2 - \sum_{i=0}^{N} |V_{\rm i}|^2
  \label{eqn:pa-ia}
\end{equation}

when using the antenna voltages directly, or

\begin{equation}
  P_{\rm PC} = 2\times Re[\sum_{i=0}^{N-1} \sum_{j=i+1}^{N-1} V_{\rm ij} e^{-i\phi_{i\rm j}}]
  \label{eqn:pc}
\end{equation}

when using the visibilities. Here $P_{\rm PC}$ is the post-correlation beam signal, $V_{\rm i}$ and $\phi_{\rm i}$ 
are delay and fringe corrected voltages and the beam steering phase of the $i^{th}$ antenna, $V_{\rm ij}$ and 
$\phi_{\rm ij}$ are the raw visibility and beam steering phase for the baseline between the $i^{th}$ and 
$j^{th}$ antenna. We also show in Fig.~\ref{fig:PA-IA_block}, schematic block diagrams of these two ways of 
forming the post-correlation beams. In section~\ref{sec:compute} we compare the computational costs of 
these two forms of beamformation.

\section{Comparison of different beamformation schemes}
\label{sec:compare}

We use a number of data sets to compare these different beamformation schemes. The first set of data is based on 
{\it SIGPROC}\footnote{\url{https://github.com/SixByNine/sigproc/}} {\it filterbank} data from uGMRT GWB backend  observations in the 300-500~MHz band of the the slow ($P \sim 8.5$~seconds)  pulsar PSR~J2144$-$3933.  Data from the IA and PA beams formed in  real time using the GWB were recorded, and the PC beam was formed offline using the difference between the 
PA and IA beams as described above (eqn.~\ref{eqn:pa-ia}). 
Simulated pulsar signals were injected into the IA and PA {\it filterbank} data files using the {\it inject\_pulsar} 
routine of {\it SIGPROC} pulsar package. A total of 12 data sets (each for IA, PA and PA$-$IA) were generated in this way, where the 
difference between the data sets is the period of the injected pulsar signal, this varies from 25~ms to 128~seconds. 
The original data also of course contains the signal for PSR~2144$-$3933, making for a total of 13 data sets from the uGMRT data. 
In addition we also used data from GMRT GSB backend observations at 607~MHz of PSR~J2144$-$3933. The Nyquist sampled antenna voltages were recorded on disk, and all beamformation as well 
as correlation was done offline. These data sets allow us to compare the performance of the different beamforming 
schemes with the exact same input data. The GSB data set also allows us to compare the two different ways of 
post-correlation beamformation discussed above.

\begin{figure}[htb]
        \includegraphics[width=2.1in,angle=0]{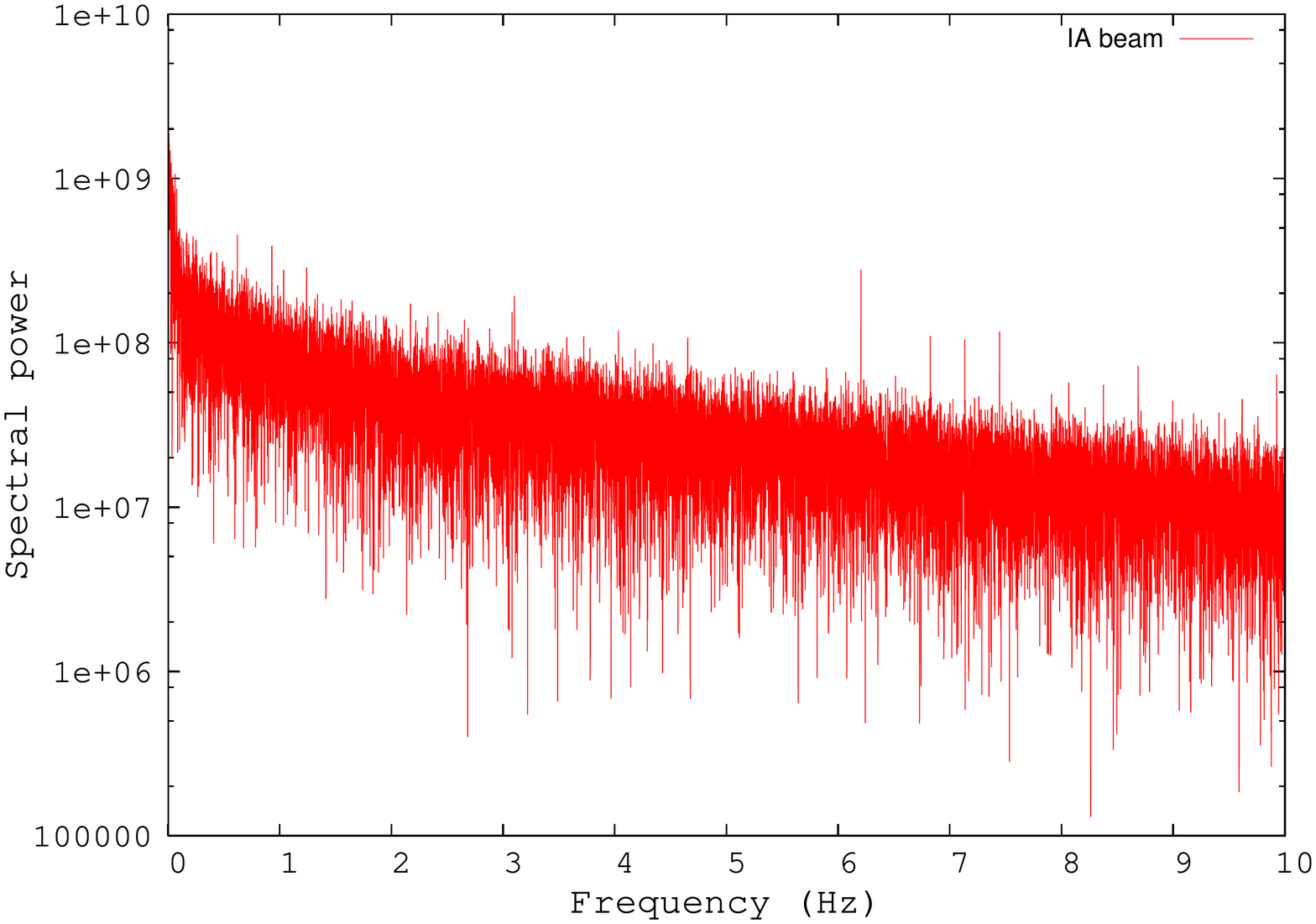}
        \includegraphics[width=2.1in,angle=0]{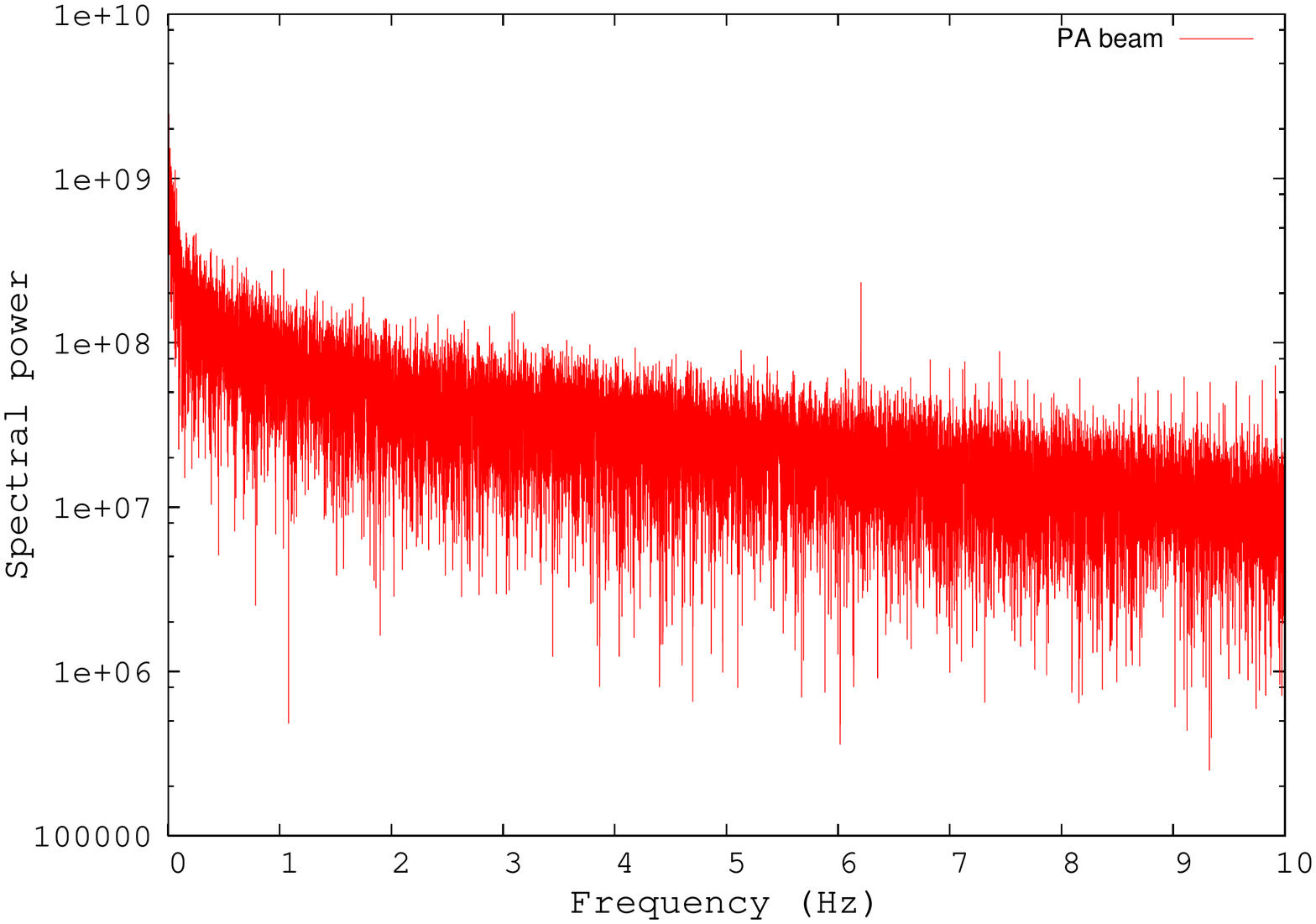}
        \includegraphics[width=2.1in,angle=0]{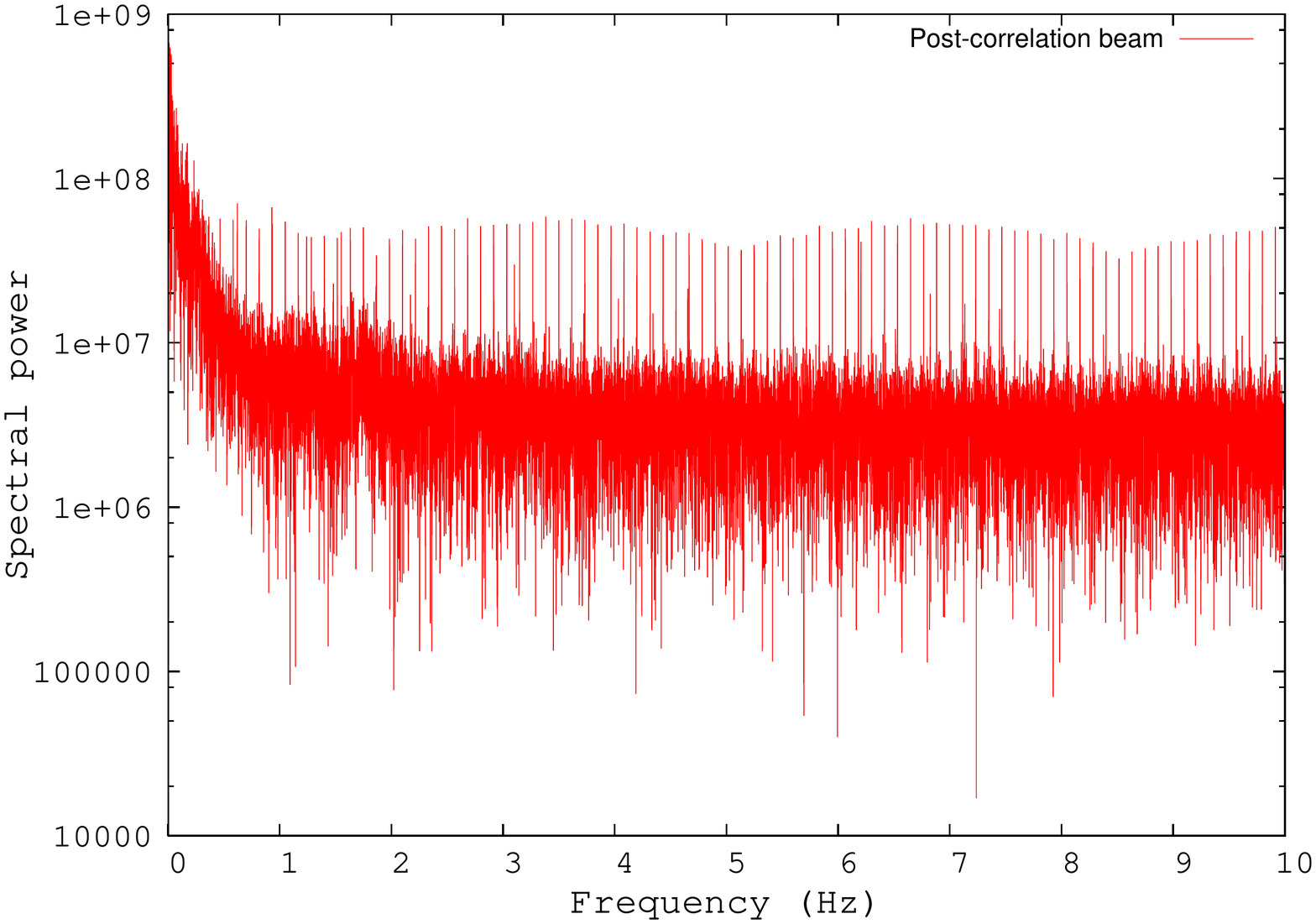}
    \caption{Power spectra for the IA, PA and PC beams for PSR J2144$-$3933.
The harmonics of the pulsar are spaced at 0.12 Hz. As can be seen, the power spectra
for the PA and IA beams are essentially the same. However, there is
an order of magnitude reduction in red noise for the PC beam. This enables the detection of
low order harmonics which are completely buried in the noise for the IA and PA beams. See the text for more details}
    \label{fig:power_spectrum}
\end{figure}

\begin{figure}
        \includegraphics[width=4in,angle=0]{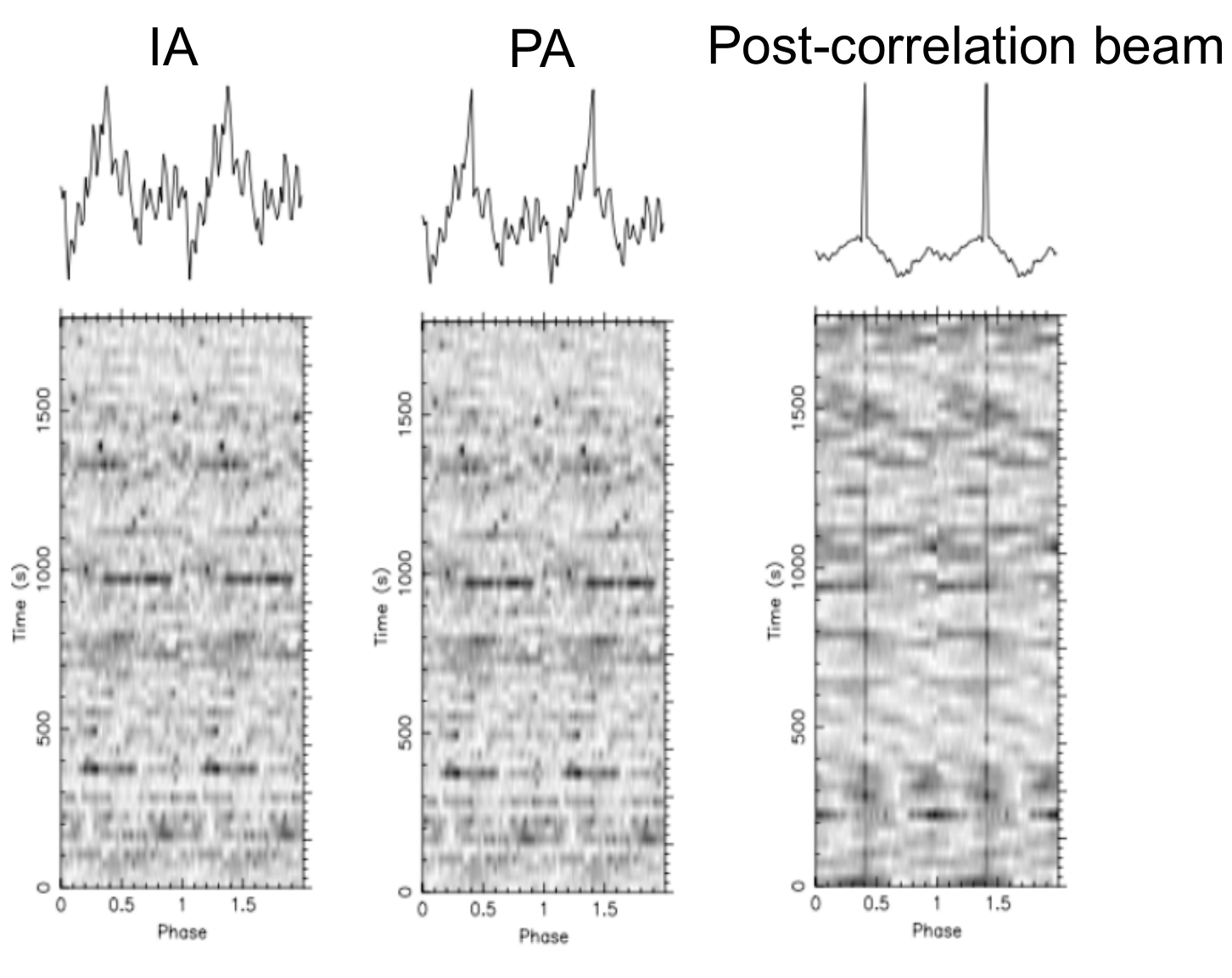}
    \caption{Folded profiles of PSR J2144$-$3933 observed with the uGMRT 300$-$500~MHz bands.}
    \label{fig:folded_profile}
\end{figure}

In Fig.~\ref{fig:power_spectrum} we show the low frequency end of the power spectra after de-dispersion 
for the three different beamforming modes using the uGMRT data. As can be seen, the power spectra for the 
PA and IA beams are essentially the same, since, as mentioned above, this part of the power spectrum is dominated  by the instrumental red noise and the RFI that is contained in the auto-correlation spectra. Consistent with 
this, the PC beam, which does not contain auto-correlation data has significantly lower noise. This `de-reddening' 
of the power spectrum should greatly ease the problem of detecting slow pulsars. Indeed, one can see that in 
the PC beam, the signal from the 8.5~seconds pulsar is detectable from the 1st harmonic onwards. For the IA or PA beam 
on the other hand, only harmonics beyond the $\sim$60th harmonic are visible in the power spectrum.
Fig. \ref{fig:folded_profile} shows the folded profiles of PSR J2144$-$3933 for these  IA, PA and PC beam data. 
A systematic and significant improvement in the signal to noise ratio is clearly visible even to the eye as one goes 
from the IA beam to the PA beam and PC beam. The PC beam's SNR is $\sim 5 - 6$ times better than that of the PA beam. 
This clearly shows the dramatic improvement in the detectability of slow pulsars when the noise and systematics contained 
in the auto-correlation spectra is eliminated. We note that the beams were formed using all of the input data, i.e. 
there has been no effort at RFI mitigation. We discuss below specific advantages the PC beam offers as far as 
targeted removal of RFI is concerned.

\begin{figure}
        \includegraphics[width=4in,angle=0]{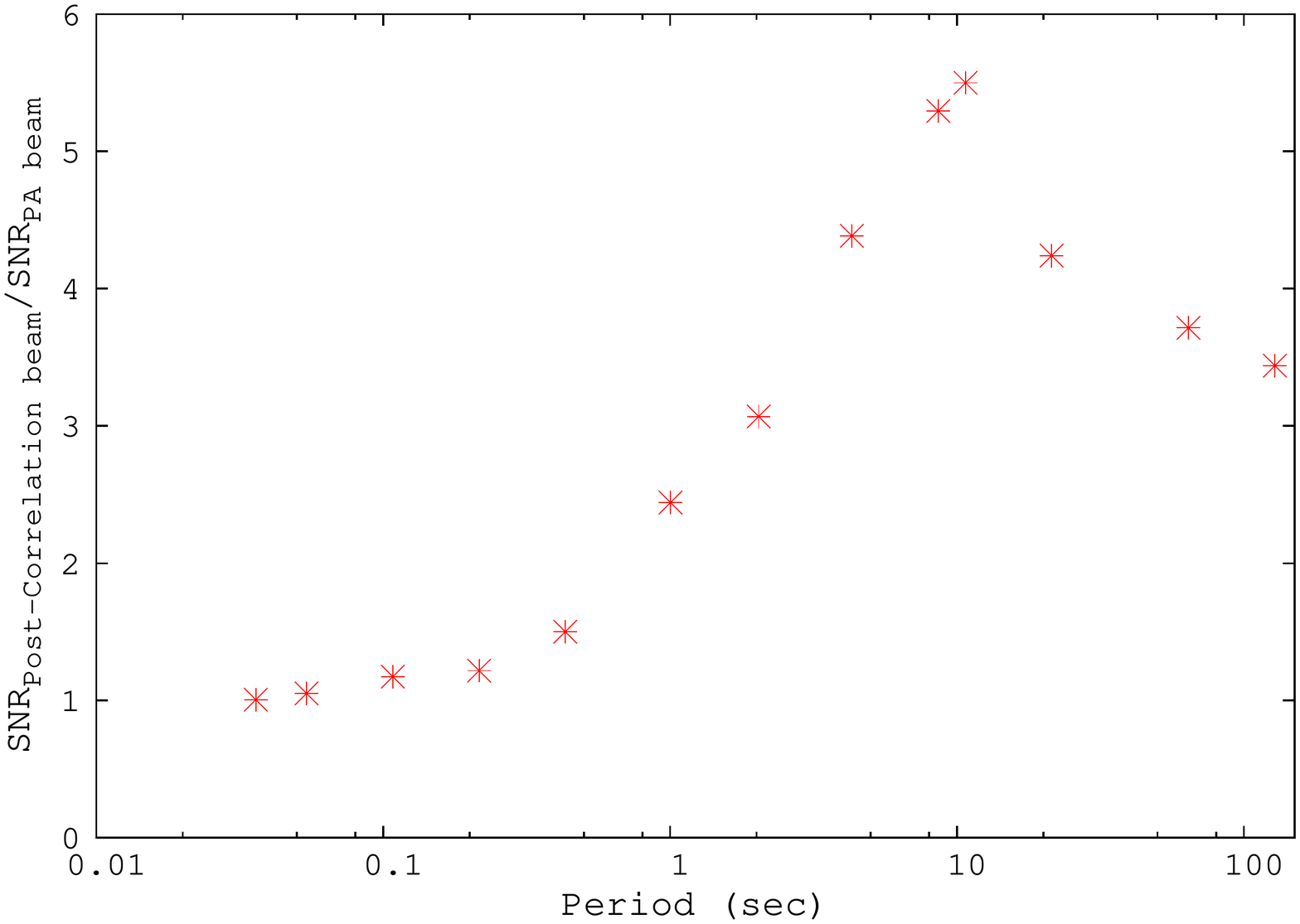}
    \caption{Sensitivity improvement with post-correlation beam compared to the PA beam as function of pulse period.}
    \label{fig:sensitivity_improvement}
\end{figure}

In Fig. \ref{fig:sensitivity_improvement} we show the ratio of the SNRs of the PC and PA beams as a function of the pulse period. 
This plot was generated using the data for the simulated pulsars as well as the data for PSR~J2144$-$3933. In all cases the 
PC beam has a higher SNR than the PA beam. The PC beam SNR is about 10\% better than that of the PA beam for a spin period 
of 25~ms; this difference reaches factors of 5$-$6 for spin periods of $\sim 10$~seconds. Beyond spin periods of $\sim 10$~seconds, 
the increase in the SNR is not as large, but it is still as much as a factor of $\sim 3$ for spin periods as long as 100~seconds, 
(i.e. spin frequency $\sim$ 0.01~Hz). This is due to the fact that red noise in PC beam also goes up below 0.1 Hz, as can 
be seen in Fig.~\ref{fig:power_spectrum}.

\begin{figure}
        \includegraphics[width=4.5in,angle=0]{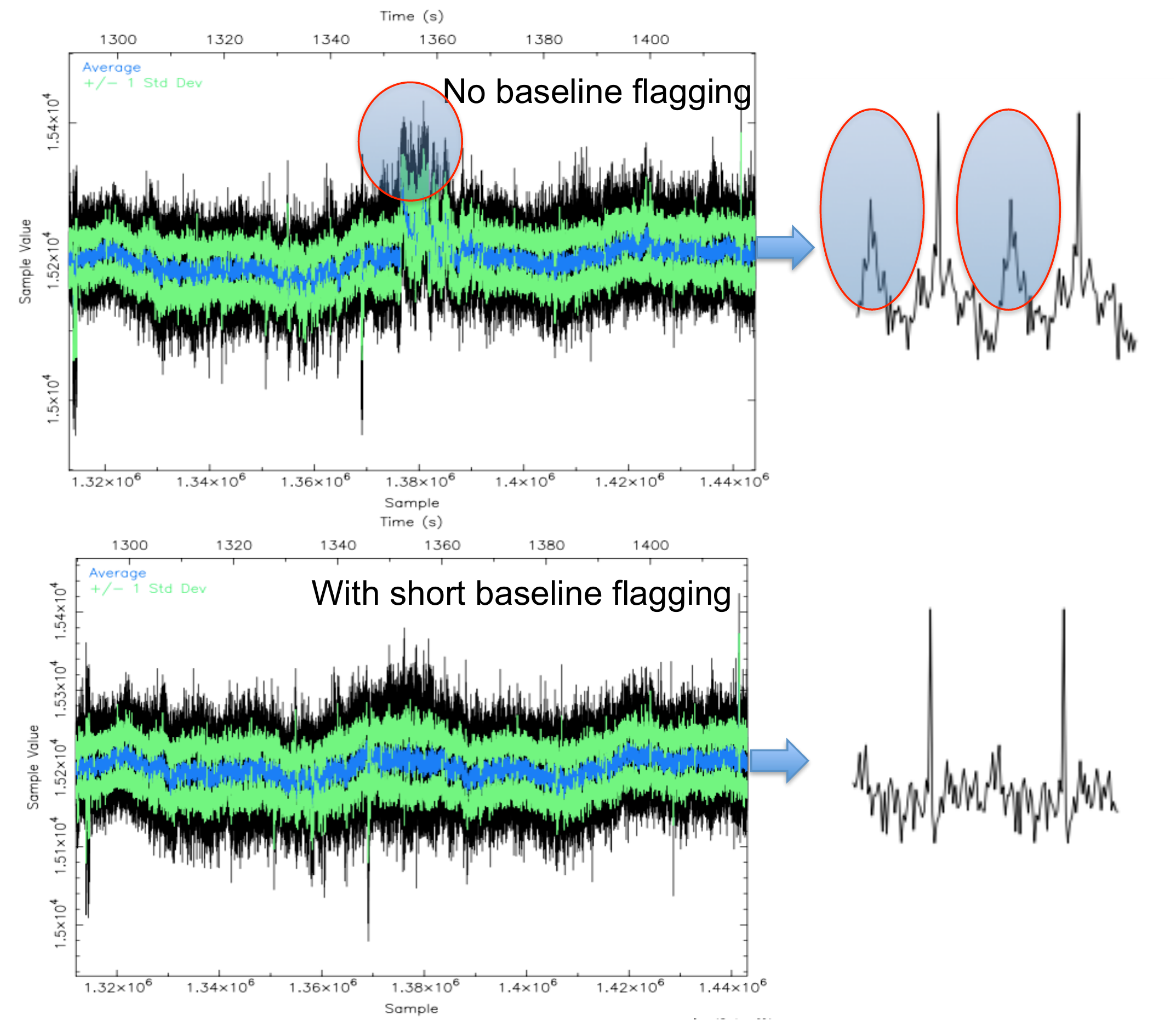}
    \caption{Visibility based post-correlation beamformation. The upper panel shows the time series when one uses all the data, while the lower panel shows the time series when the short baselines (which tend to be those most affected by RFI) are flagged. The plots on the side show the corresponding folded profiles. A dramatic decrease in the systematics can be seen when one flags the baselines with RFI. See the text for more details.}
    \label{fig:baseline_flag}
\end{figure}

The two methods of post-correlation beamformation presented in  Sec.~\ref{sec:postcorr} are mathematically equivalent. 
One might imagine then that all that distinguishes these two methods is their respective computational costs. We discuss 
this issue in Sec.~\ref{sec:compute} below. However, there is one further way in which these two methods are different, viz. in the possibilities that they offer for identifying and removing RFI. When the PC beam is formed as the PA$-$IA beam, 
one can only flag out data at the granularity of an antenna. When forming the PC beam from the visibilities, one can flag out 
data at the granularity of baselines. This is particularly useful in arrays which contain antennas at a range of separations. 
Often data from the short baselines contain significantly more RFI than the data from long baselines. Since nearby antennas 
also have baselines with more distant antennas, this could allow one to greatly eliminate the RFI while retaining much of 
the raw sensitivity. We show in Fig. \ref{fig:baseline_flag} that short time-scale (i.e. few seconds) RFI bursts present 
in the PC (visibility based) beam can be removed by flagging out the data from all the baselines shorter than a $\sim~$450 metres 
 (i.e. 3\% to 4\% of the total GMRT baselines). As shown in the figure, these RFI 
burst generate pseudo pulse like features in the folded profile of PSR J2144$-$3933; flagging the short baseline very 
effectively mitigates the problem. We note that the flagging done here was ``blind'', i.e. short baselines were flagged, 
without looking at the data quality on these baselines. In principle one could use flagging algorithms 
(for e.g. such as FLAGCAL \cite{Prasad12}) to automatically identify and flag only those baselines which actually do have RFI.
  
\begin{figure}
        \includegraphics[width=4in,angle=0]{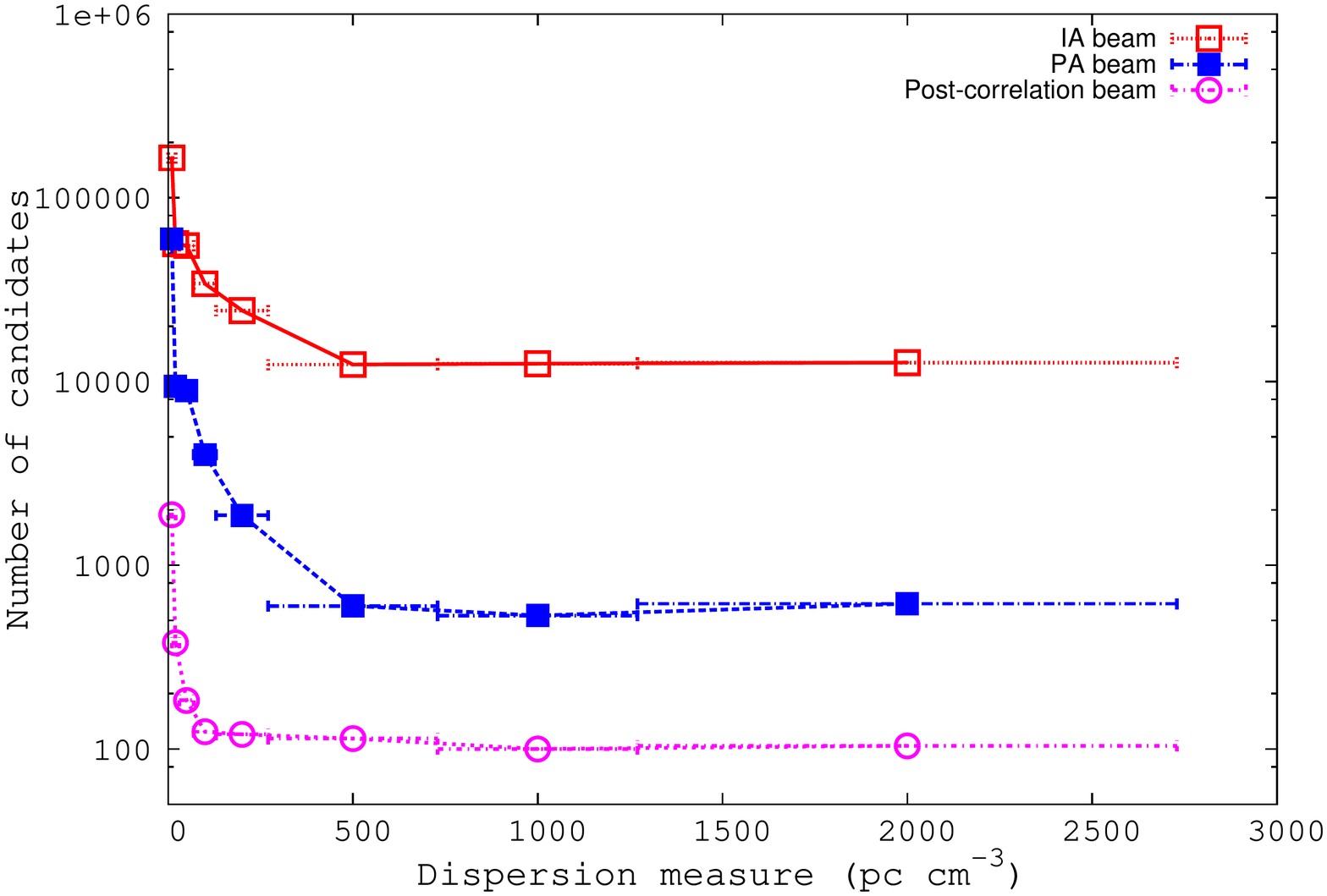}
    \caption{Number of candidate detected from IA, PA and post-correlation beam for simulated FRB events with
various DMs for a data of 60~seconds duration. There are 8 FRB events simulated at DM of 10, 20, 50, 100, 200, 500, 
1000 and 2000 pc $cm^{-3}$. The DM range used in each steps for searches are shown by horizontal bars around each central DMs.
This detection rate is for a threshold of 5$\sigma$. The number
of (false) detections from the IA beam is almost two order of magnitude higher than the post-correlation beam even
at a DM of 2000 pc $cm^{-3}$.}
    \label{fig:FRB_rate}
\end{figure}

So far we have been comparing the characteristics of the IA, PA and PC beams as far as detecting pulsars are concerned. 
Another class of pulsed signals that is of great interest currently are transients such as FRBs which emit single pulses. 
While observing with an interferometric array, one can save the visibilities for candidate events, so that one could also 
image the field in order to localize any confirmed sources (see e.g. \cite{Bhat13}). As discussed in detail in \cite{Bhat13}, 
in such searches, it is important to reduce false positives as much as possible, in order to minimize the amount of data 
that has to be saved and processed. Since the PC beam contains far less RFI than the IA and PA beam, one would expect that 
the number of false positives in the PC beam would also be less than that for the other beamforming modes. 
Fig.~\ref{fig:FRB_rate} shows the number of candidates detected from IA, PA and PC beam for simulated FRB events with 
various DMs injected in the same uGMRT 300$-$500~MHz band data as discussed above. The signals were injected using the 
same {\it inject\_pulsar} routine, but with pulse period being much larger than the duration of the data (i.e. 60~seconds). 
The post-correlation beam is formed as PA$-$IA. There are 8 FRB events simulated at DM of 10, 20, 50, 100, 200, 500, 
1000 and 2000 pc $cm^{-3}$. {\it PRESTO} based single pulse search were performed for all these three beams 
over a range of DMs (indicated by the error-bars). As can be seen from the figure, the number of triggers from the
PC beam is almost two orders of magnitude lower than that the IA beam even at a DM as high as 2000 pc cm$^{-3}$. 
The number of false positives in the PC beam data is also a factor of $\sim 5$ less than that found in the PA beam data.  
Interestingly, over the full FRB DM search space (i.e. 250 to 2600 pc cm$^{-3}$), the candidate detection rate is almost 
constant for the PC beam. The percentage of true positives at the highest (2000 pc $cm^{-3}$) 
DM value of the simulations for IA, PA and PC beams are 0.0004, 0.8 and 5 respectively.
We note that this plot was generated for candidates detected above a threshold of 
5$\sigma$ in order to make the uGMRT 300$-$500~MHz PC beam sensitive enough to detect all the known FRBs 
(ignoring frequency dependent scattering and spectral steepening). At this threshold the uGMRT IA beam detects only 
30\% of the known FRBs. Raising the threshold to 10$\sigma$, generates only very few triggers from the PC beam for this data, 
whereas IA beam continues to be equally corrupted. The recently detected FRBs are all at the lower end of the FRB flux distribution, 
all of these will be completely missed at the sensitivity offered by the IA beam. As is the case for FRBs, the PC beam data 
would also contain far fewer false positives in searches for other transients such as RRATs (most of which have 
DM $<$ 300 pc cm$^{-3}$). Both manual as well as automated searches for RRATs in the IA beam data would be swamped by 
the large number of false positives. Ways of overcoming this problem by forming multiple incoherent sub-array beams and using 
co-incidence filtering are discussed in \cite{Bhat13}. However splitting antennas in sub-arrays significantly reduces 
the survey sensitivity. Another major difference between the IA and PC beam is of course the field of view. In blind surveys 
one would like to have as large a field of view as possible, in which case PC beamformation is not competitive, unless one is 
able to form multiple beams. In the next section we detail the computational cost involved in forming multiple PC beams.

\section{Computational requirements}
\label{sec:compute}

As discussed above, there are two different ways of forming the post-correlation (PC)  beam. The first is 
via the difference of the PA and IA beams, while the second is via a phased addition of the visibilities. 
While the PC beamformed in these two ways is mathematically equivalent, we also saw that operationally 
the visibility route might have some advantage because of the better opportunities it provides for 
flagging data affected by RFI. Here we take a look at the difference in the amount of computation 
required to make the PC beam in these two ways. To start with, we note that correlators require 
a fan out of the data, i.e. in order to correlate the data from one antenna with all other antennas 
one needs multiple copies of the data stream. On the other hand beamformation operates on the data stream 
from each antenna independently, except in the final addition stage. This would lead to differences in 
architecture. Here we do not look at this in detail, but instead focus only on the number of computations 
required to make the PC beam in these two different ways.

We start by defining the parameters needed to determine the required computation, with the assumed 
value of the parameter for the GMRT (where relevant) given in parenthesis. Let the total number of 
beams to be formed (each with an independent phase center) be $N_B$.  The total number of antennas 
is  $N_{a}$ (30), the bandwidth of operation $B$ (200~MHz), the number of time samples in a given 
FFT block $N_{f}$ (4096 for 2048 spectral channels) and the number of FFT blocks per integration $N_{b}$. In terms of these 
parameters the channel resolution is $\Delta{\nu}$ = $\frac{B}{N_{f}}$ and the 
integration time $\Delta{\tau}$ = $\frac{N_{b}N_{f}}{2B}$.

The total computational load (in number of operations per second) for PA$-$IA beamformation for one integration is
\begin{equation}
5N_{a}N_{b}N_{f}logN_{f} + N_{B}N_{a}N_{b}N_{f} + N_{B}N_{f}(N_{a}N_{b}+N_{b}-1) + N_{b}N_{a}N_{f} + (N_{a}-1)(N_{b}-1)N_{f} + N_{B}N_{f}
\end{equation}
and consists of the following components:
\begin{itemize}
  \item 5$N_{a}N_{b}N_{f}logN_{f}$ for FFT
  \item $N_{B}N_{a}N_{b}N_{f}$ for fringe and fractional-delay corrections as well as beam steering
  \item $N_{B}N_{f}(N_{a}N_{b}+N_{b}-1)$ for PA beamformation including addition, squaring and integration
  \item $N_{b}N_{a}N_{f}$ + $(N_{a}-1)(N_{b}-1)N_{f}$ for IA beamformation including squaring, addition and integration
  \item $N_{B}N_{f}$ for the PA-IA operation
\end{itemize}

We note that for PA$-$IA beamformation the phase corrections for beam steering needs to done before 
antenna addition, which requires working at the FFT resolution. However, since the maximum fringe rate 
of the GMRT is $\pm$ 5 Hz \citep{Chengalur98}, the maximum possible delay change even over a period as 
large as 1~ms is much smaller than the Nyquist sampling resolution. This means that for PC beam  formation from 
the visibilities we can do the differential beam steering after the visibilities have been computed. 
The total computation load (in number of operations per second) for visibility based PC beamformation for one integration is
\begin{equation}
\begin{split}
5N_{a}N_{b}N_{f}logN_{f} + N_{a}N_{b}N_{f} + N_{b}N_{f}\frac{N_{a}(N_{a}-1)}{2} + (N_{b}-1)N_{f}\frac{N_{a}(N_{a}-1)}{2} \\
+ N_{B}N_{f}\frac{N_{a}(N_{a}-1)}{2} + N_{B}N_{f}[\frac{N_{a}(N_{a}-1)}{2}-1]
\end{split}
\end{equation}
and  consists of the following components:
\begin{itemize}
\item 5$N_{a}N_{b}N_{f}logN_{f}$ for FFT
\item  $N_{a}N_{b}N_{f}$ for fringe and fractional-delay corrections at the pointing centre (common to all beams)
\item $N_{b}N_{f}\frac{N_{a}(N_{a}-1)}{2}$ + $(N_{b}-1)N_{f}\frac{N_{a}(N_{a}-1)}{2}$ for correlation including multiplications
  and additions
\item $N_{B}N_{f}\frac{N_{a}(N_{a}-1)}{2}$ for phase corrections required for steering the individual beams
\item $N_{B}N_{f}[\frac{N_{a}(N_{a}-1)}{2}-1]$ for visibility addition for the beamformation
\end{itemize}

For the given GMRT configurations with 1600 beams, in PA$-$IA based PC beamformation (i.e. Eq 7), 
term 2 (fringe, fractional delay and beam steering) and 3 (PA) dominates equally and at least 20-times 
higher than any other terms. Whereas for visibility based PC beamformation (i.e. Eq. 8), 
term 4 (beam steering) and 5 (visibility addition) dominates, but they are only an order of magnitude 
higher than the next dominating term 2 (FFT).  However the contributions of term 4 and 5 of Eq. 8 increase 
by many-folds than any other terms as beams are formed at high time resolution.
Considering these, one would expect that PA$-$IA beamformation is computationally cheaper for a small number 
of high time resolution beams, while the visibility based beamformation would be computationally cheaper for 
a large number of beams at low time resolution. The cross over point would depend also on the total number of elements. 
We show in Fig.~\ref{fig:compute_cost}  a comparison of these two computational loads as a function of the total number of beams formed and 
the time resolution for a GMRT like array of 30 antennas (upper panel) as well as an SKA Phase1 Mid like array of 256 antennas. 
For the GMRT array we use 200~MHz instantaneous bandwidth with 2048 spectral channels at 163.84~$\mu$s (upper right) and 1.31~ms (upper left) 
time resolution. For SKA Phase1 Mid array \citep{Levin17} we use 300~MHz instantaneous bandwidth with 4096 spectral 
channels at 64~$\mu$s (lower right) and 2.048~ms (lower left) time resolution. The figures clearly bring out the 
broad trends expected with time resolution and number of beams. For the GMRT, visibility beamformation is economical 
compared to PA$-$IA beamformation for time resolutions $\geq$ 163.84~$\mu$s and for $\geq$ 10 beams. For SKA Phase1 Mid array, 
visibility beamformation is economical compared to PA$-$IA beamformation for time resolutions $\geq$ 2.048~ms and 
for $\geq$ 800 beams. A configuration with a small number of high-time resolution beams would be useful in searches 
for pulsars (specially MSPs) via targeted observations of globular clusters (GCs). GCs are the most likely hosts 
of exotic binary systems, like MSP-main sequence binaries, highly eccentric binaries, MSPs in evolutionary phases 
like Redback, Black Widow and MSP-Black Hole binaries, which may not form via normal stellar evolution in the disk. 
The multiple beams should be sufficient to cover the expected sky area within which MSPs expelled from the centre 
but which are still within the cluster tidal radius. A moderate number of high time-resolution beams offers an 
opportunity to greatly increase the pulsar timing efficiency in arrays where the individual elements have a 
large field of view, by allowing simultaneous observations of multiple pulsars \citep{Stappers18}. A large number of lower time 
resolution beams (as would be cheaper via the visibility route) would be useful in blind searches for all 
but the fastest pulsars.

\section{Case study of a proposed GMRT survey}
\label{sec:discussion}

The improvements seen in time-domain processing using post-correlation beamformation aided with the enhanced sensitivity 
of the uGMRT for the GHRSS phase-2 survey, provide the motivation to develop a time-domain 
survey with a post-correlation beamformer. We compute here the estimated parameters for such a survey. To benchmark, 
we consider the uGMRT 300$-$500~MHz band with 30 antennas, 200~MHz bandwidth, 2048 spectral channels, visibility 
beamformation at 1~ms time-resolution with about 128 beams covering $\sim$ 10\arcmin~FoV. We note that covering 
the entire field of view with PC beams would require $\sim$ 1600 beams.  
We estimate a survey sensitivity of $\sim$ 0.1~mJy at 400~MHz (considering the radiometer equation \ref{eqn2} 
with a 2\% loss for ignoring the auto-correlation power), for a 10$\sigma$ detection for a 10\% duty cycle, a 
post-correlation beam gain of 7 K/Jy for 200~MHz bandwidth, 10~mins of dwell time, and a system temperature of 106~K. 
We also calculate a sensitivity of 0.05~Jy as the 5$\sigma$ detection limit for 5~ms transient millisecond bursts, 
which would correspond to weak scattering \citep{Thornton13}.

\begin{figure}
        \includegraphics[width=4in,angle=0]{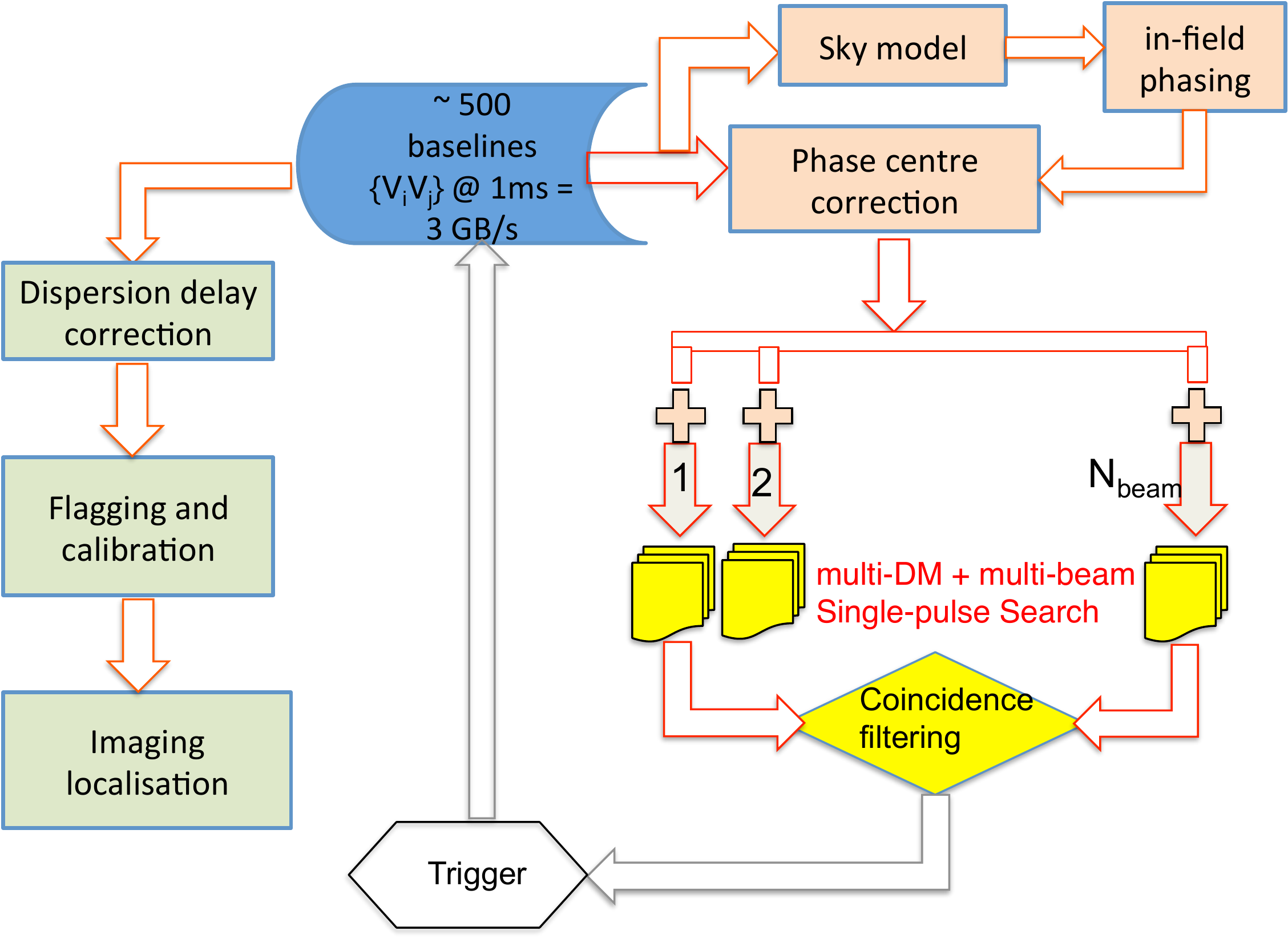}
    \caption{Proposed multi-beam time-domain survey with visibility beamformation. There are
4 functional modules each colored differently running on different compute hardware. Please refer
to the text for further details.}
    \label{fig:FRB_survey}
\end{figure}

Fig. \ref{fig:FRB_survey} shows the components required for such time-domain survey with the post-correlation 
beamformer as specified above. The required components are shown in four different colours. Visibilities computed in 
the uGMRT  backend (GWB; marked in blue) at 1~ms time resolution are transferred to the the post-correlation beamformer 
nodes (marked in orange) with an aggregate data rate of $\sim$ 3 GB/s. We aim to implement in-field phasing 
(\cite{Kudale17}) using a sky model derived from the time-averaged visibilities in order to improve the coherence 
in phasing up to baseline length of several kilometers. This optimises the GMRT phased array sensitivity beyond central compact core (most current phased array observations use only the antennas in the central square). In addition of deriving the phasing model, 
a baseline based flag masking the bad baselines will also be generated in real-time from these time-averaged 
visibilities. Coherent additions of these visibilities will result in 128 of such visibility beams. The multi-DM 
search for single pulses (colored in yellow) on each of these visibility beams would need to be executed on a 
separate FRB cluster followed by coincidence filtering to remove spurious events \citep{Bhat13}. It is also 
proposed to record these 128 beams with 1~ms time-resolution giving a total data rate of 200 MB/s into disk 
for quasi-realtime search for pulsars using the same cluster. We note that the proposed 1~ms time-resolution is 
sufficient to detect double neutron star systems, young pulsars, normal pulsars as well as object like radio magnetars. 
Visibility buffers corresponding to candidate single pulse events will be recorded at 1~ms time resolution covering 
the full DM sweep time-range. For a event at a DM of 2000 pc cm$^{-3}$, the total DM sweep time over 200~MHz band in 
uGMRT 300$-$500~MHz band is $\sim$ 50~seconds, which results in 40 GB buffer size on each of the post-correlation beam 
nodes. This means one can easily hold few buffers for accommodating the pipeline delay and flush them to a 
storage based on the real-time triggers. These visibilities will be processed through the processing blocks 
(marked in green) for millisecond imaging localization at quasi real-time. This block includes removal of 
dispersion delay followed by a flagging and calibration pipeline and snapshot imaging. 
We note that part of this imaging pipeline to localise time-domain events has already been demonstrated for 
the GHRSS phase-1 survey \citep{Bhattacharyya16}.

\section{Summary}
\label{sec:summary}

In this paper, we demonstrate that use of post-correlation beamformer for radio interferometric array results 
in a many-fold increase of the detection significance of time domain events compared to the conventional 
incoherent and coherent array beamformer. This increase in sensitivity is driven by the lower red-noise and 
RFI contamination of the post-correlation beam.  Post-correlation beamformation also allows one to use standard 
interferometric calibration techniques for calibrating the beam. We compare two different modes of post-correlation 
beamformation, viz. (1) PA-IA beamformation, which does not require computation of the visibilities and (2) 
visibility beamformation where the beam is formed from the computed visibilities. We also show that the 
PA-IA beamformation is computationally economical for a small number of high time resolution beams. 
At low time resolutions, the visibility based beamformation is computationally cheaper. 
Visibility based beamformation also allows for better control in
flagging/suppressing RFI. 
For multi-element feed system (e.g. Parkes multibeam system) or for phased array feed, 
the PC beam can also be used to subtract RFIs (correlated within the feed elements) from feed element response \citep{Kocz10}.
These new beamforming techniques could significantly
improve the sensitivity of time-domain studies with both existing (e.g. uGMRT,
JVLA) and upcoming (e.g. CHIME, \cite{Amiri18}; OWFA, \cite{Subrahmanya17}) 
radio interferometric arrays. As a specific example, we have
presented a proposed time-domain survey with the uGMRT.

 
\section{Acknowledgments}
We thank the computer group at GMRT and NCRA. We thank Mr. Harshavardhan Reddy of GMRT for 
insightful discussions on the data-rate issues in the GWB. We thank Mr. Sanjay Kudale of GMRT 
for trying out in-field phasing in generating post-correlation beams. 
Ue-Li Pen acknowledges NSERC for support. We also acknowledge the support of 
telescope operators during our test observations which 
also involves data intensive baseband 
recording observations. The GMRT is run by the National Centre for Radio Astrophysics of the 
Tata Institute of Fundamental Research.


  \begin{figure}[!ht]
    \subfloat[\label{fig:ia}]{%
      \includegraphics[width=4.2in,angle=0]{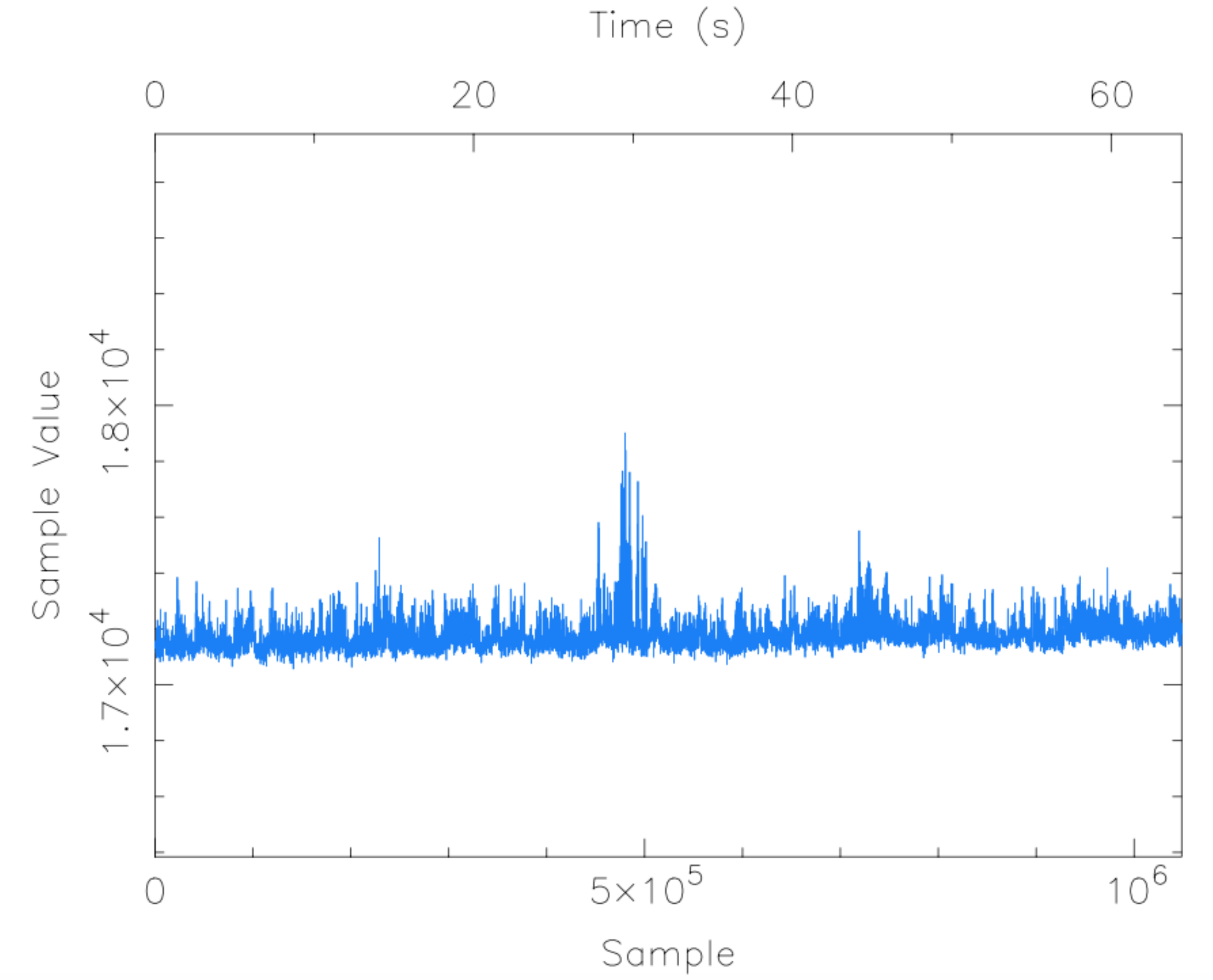}
    }\\
    \hfill
    \subfloat[\label{fig:pa}]{%
      \includegraphics[width=4.2in,angle=0]{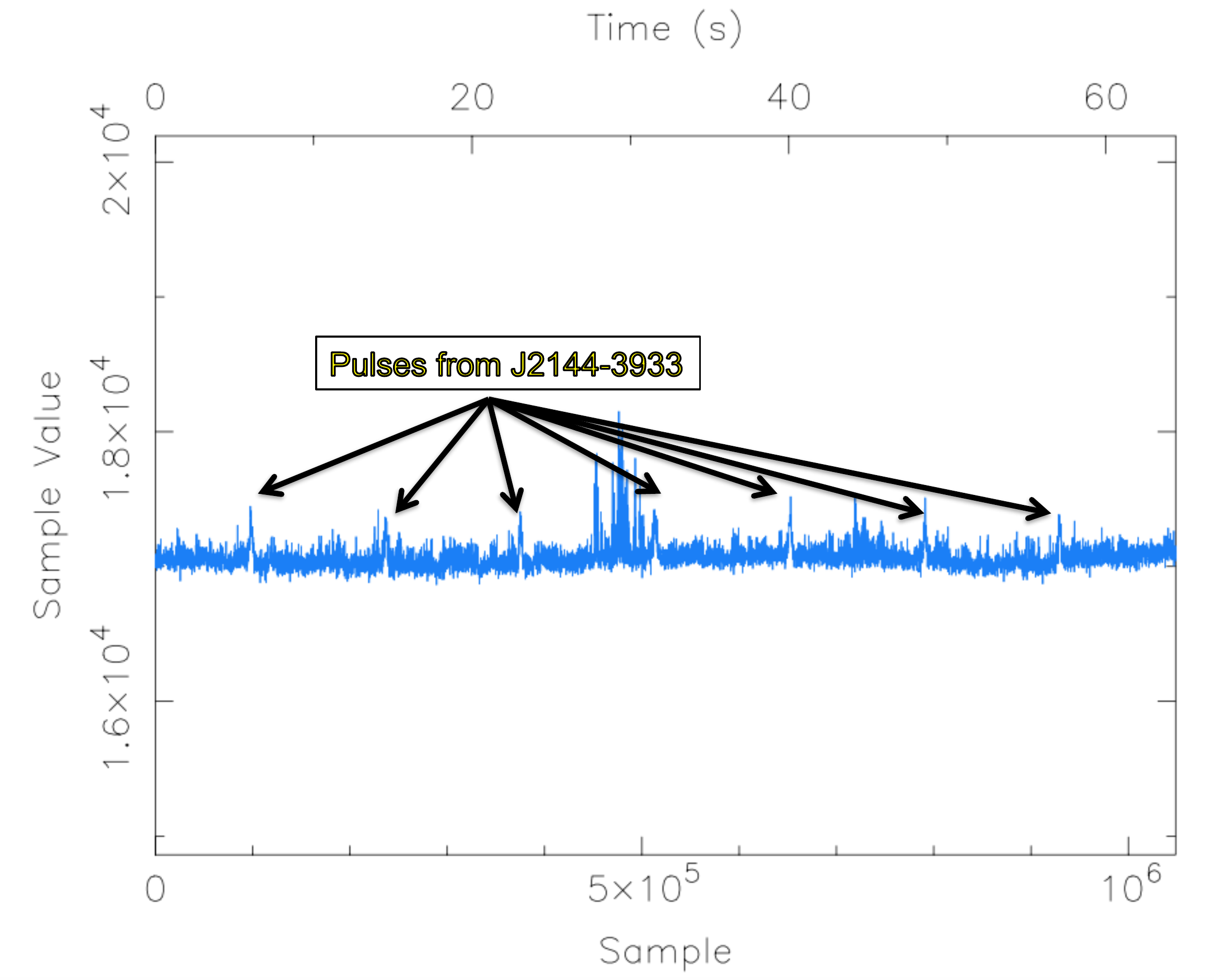}
    }\\
    \hfill
  \caption{Dedispersed time-series for PSR J2144$-$3933 from simultaneous observations of (a) IA beam, (b) PA beam are shown here.
      The plots were generated using the {\it PRESTO} software tools. The plot shows the mean value computed using moving average of
      8 time samples. The y-axis scale is different for the different panels. Fewer RFI bursts are seen in the PA beam as compared to
      the IA beam.  Also individual single pulses (as marked in the plot) are visible for PA beam while these pulses are lost
      in the noise of the IA beam.}
  \end{figure}
  \clearpage
  \begin{figure}[!ht]\ContinuedFloat
    \subfloat[\label{fig:pa-ia}]{%
      \includegraphics[width=4.2in,angle=0]{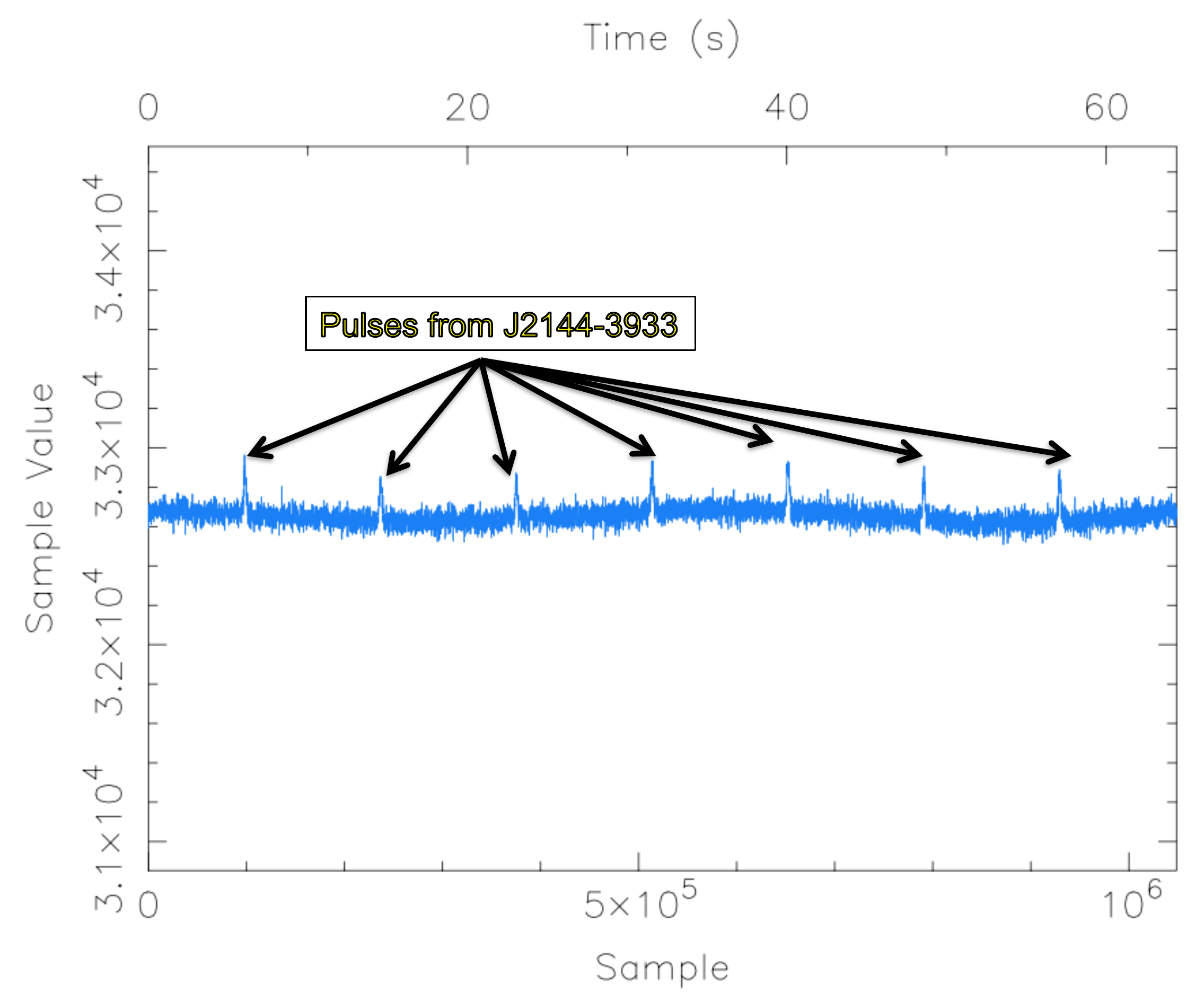}
    }
    \caption{(c) Dedispersed time-series for PSR J2144$-$3933 from the post-correlation beam. As can be seen, there is a significant
      improvement in the immunity against RFI, and the individual pulses can be clearly seen.}
    \label{fig:rednoise}
  \end{figure}
\begin{figure}[htb]
        \subfloat[\label{fig:compute_cost}]{%
        \includegraphics[width=3.5in,angle=0]{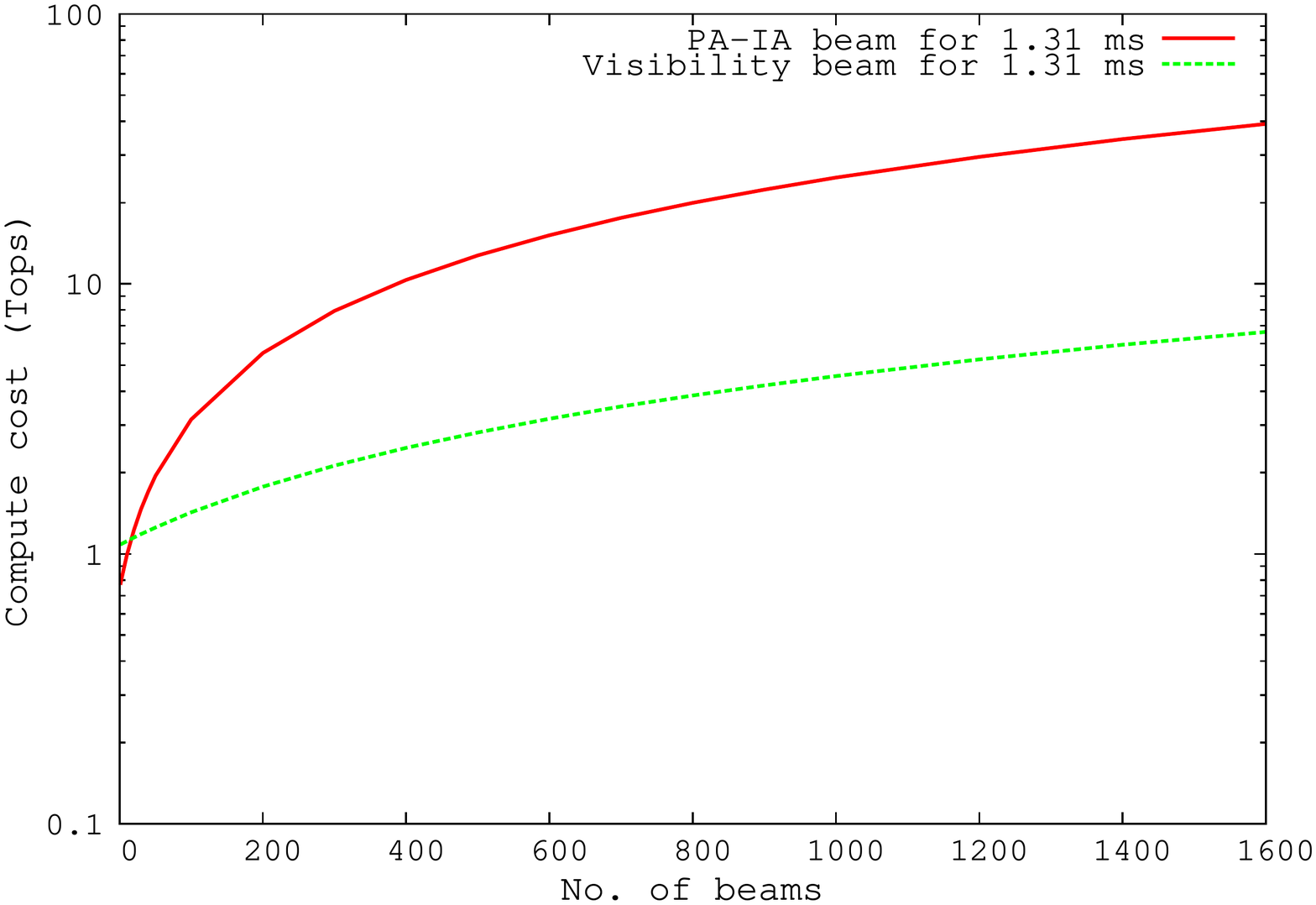}
        \includegraphics[width=3.5in,angle=0]{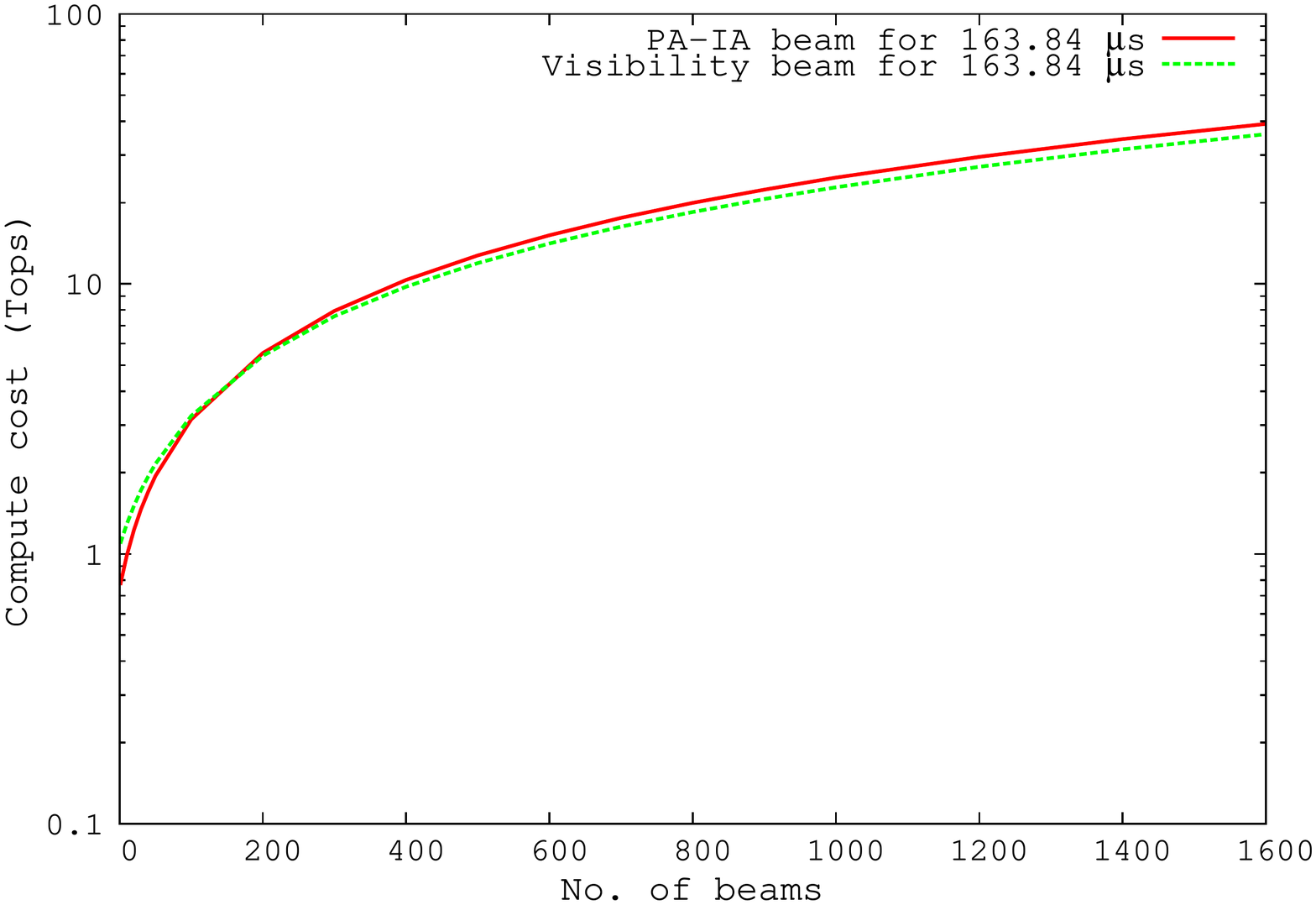}
        }
        \hfill
        \hfill
        \subfloat[\label{fig:compute_cost}]{%
        \includegraphics[width=3.5in,angle=0]{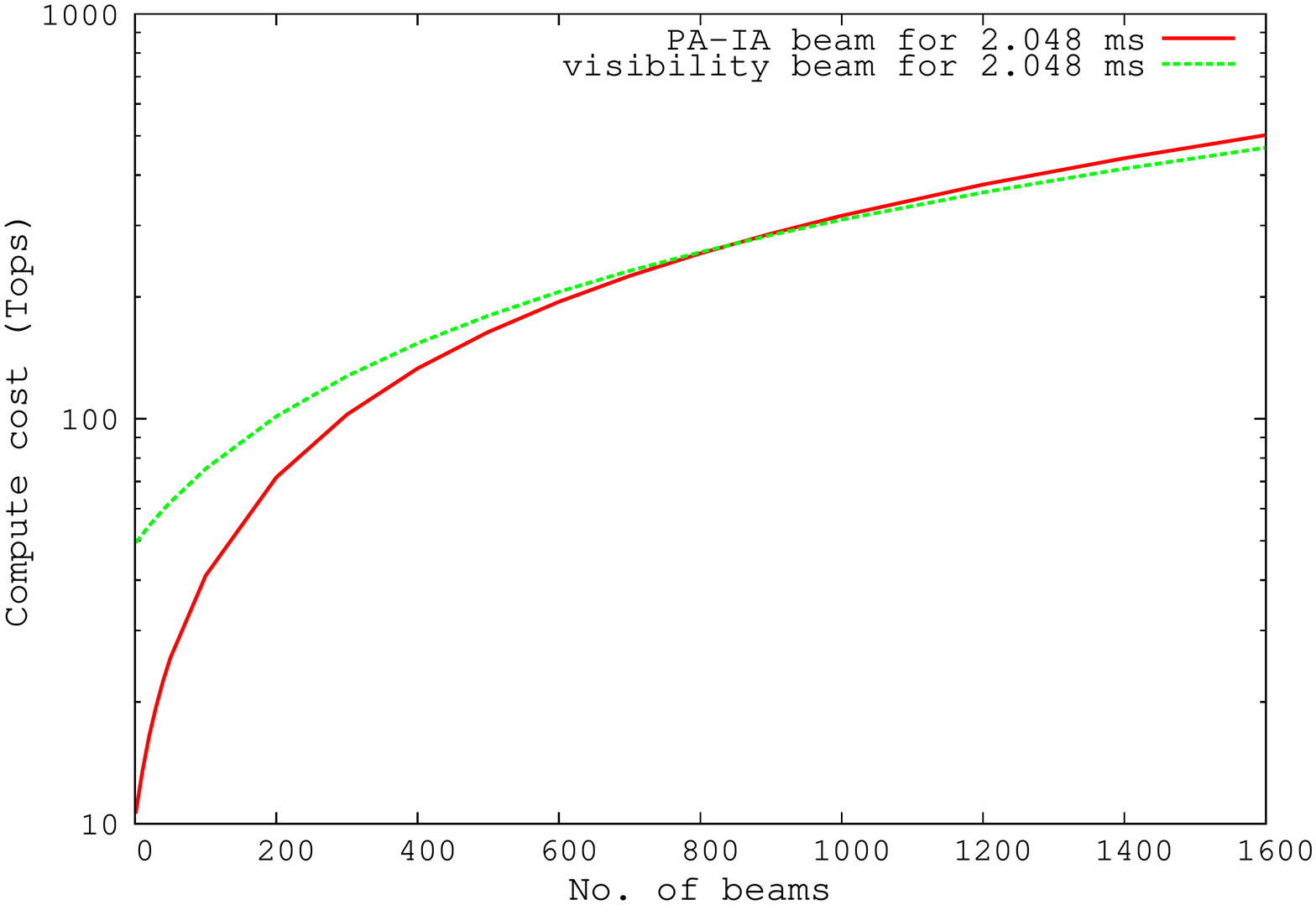}
        \includegraphics[width=3.5in,angle=0]{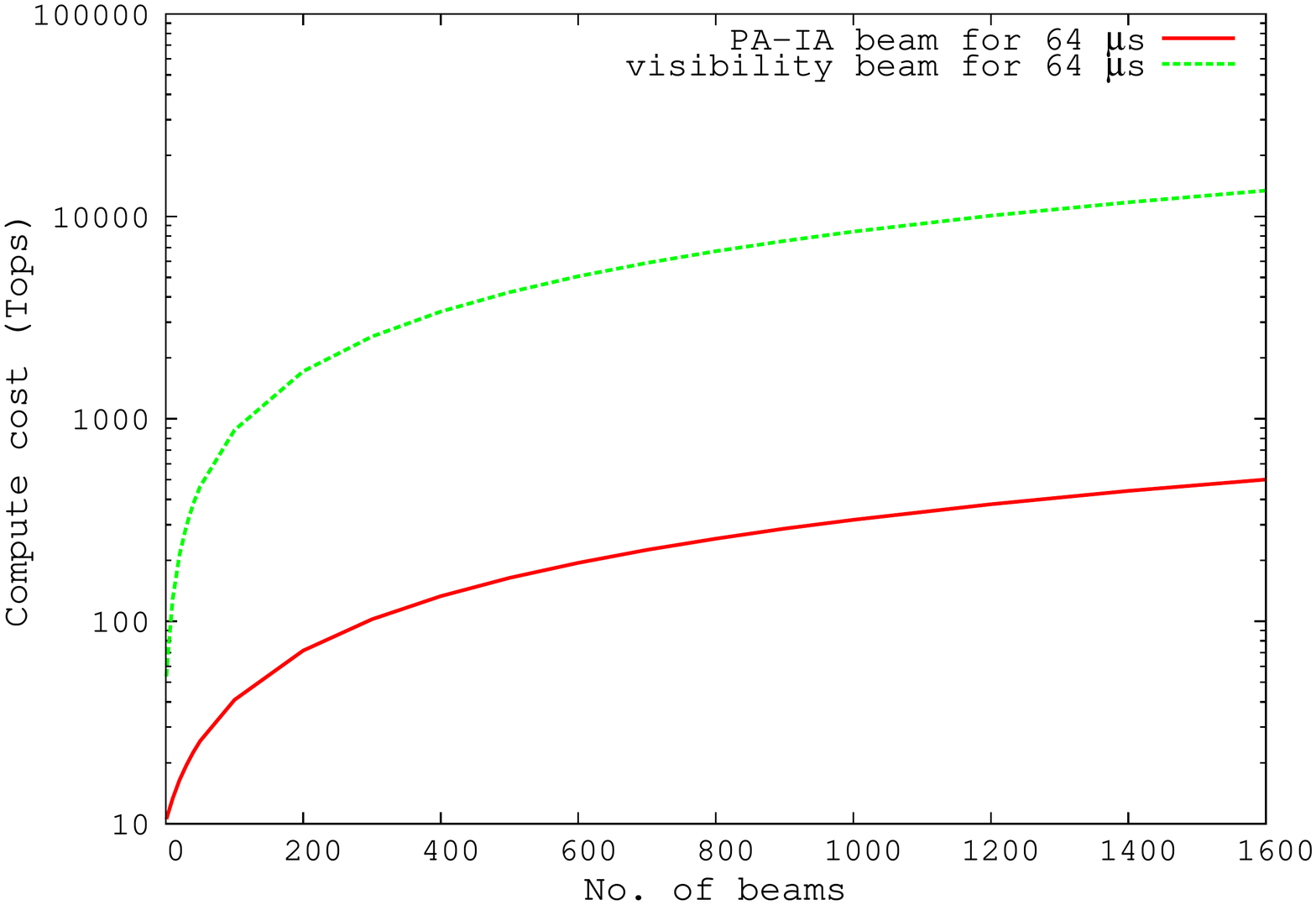}
        }
    \caption{Comparison of compute costs for two ways of forming post-correlation beams for the GMRT (a)
and the SKA Phase1 Mid array (b). For the GMRT, 1.31~ms (left panel) and 163.84~$\mu$s (right panel)
output time-resolution and for SKA Phase1 Mid, 2~ms (left panel) and 64~$\mu$s (right panel) out time-resolution
are plotted. PA-IA beams are marked in red  and visibility beams are marked in green. Cost for both the beamformation modes
are similar at 163.84~$\mu$s time-resolution for GMRT and at 2~ms time-resolution for SKA Phase1 Mid.}
    \label{fig:compute_cost}
\end{figure}

\end{document}